\documentclass[useAMS,usenatbib]{mn2e}
\usepackage{color}
\usepackage{times}
\usepackage{amsmath}
\usepackage{amssymb}
\usepackage{amsbsy}
\usepackage{graphicx}
\usepackage{subfigure}
\usepackage{paralist}

\def\kpc{\,{\rm kpc}}

\def\pc{\,{\rm pc}}
\def\kpcMyr{\,{\rm kpc\,Myr^{-1}}}

\def\kms{\,{\rm km\,s^{-1}}}

\def\Gyr{\,{\rm Gyr}}

\def\ActionUnits{\,{\rm kpc^{2}Myr^{-1}}}
\def\percent{\text{ per cent}}

\title[Dangers of fitting orbits to streams]{Stream-orbit misalignment I: The dangers of orbit-fitting}
\author[J. L. Sanders and J. J. Binney]{Jason L. Sanders$^1$\thanks{E-mail: jason.sanders@physics.ox.ac.uk} and James Binney$^1$\\
$^1$ Rudolf Peierls Centre for Theoretical Physics, Keble Road, Oxford OX1
3NP, UK\\ 
}

\pagerange{\pageref{firstpage}--\pageref{lastpage}} \pubyear{2013}
\label{firstpage}

\begin{document}
\maketitle

\begin{abstract}
Tidal streams don't, in general, delineate orbits. A stream-orbit misalignment is expected to lead to biases when using orbit-fitting to constrain models for the Galactic potential. In this first of two papers we discuss the expected magnitude of the misalignment and the resulting dangers of using orbit-fitting algorithms to constrain the potential. We summarize data for known streams which should prove useful for constraining the Galactic potential, and compute their actions in a realistic Galactic potential. We go on to discuss the formation of tidal streams in angle-action space, and explain why, in general, streams do not delineate orbits. The magnitude of the stream-orbit misalignment is quantified for a logarithmic potential and a multi-component Galactic potential. Specifically, we focus on the expected misalignment for the known streams. By introducing a two-parameter family of realistic Galactic potentials we demonstrate that assuming these streams delineate orbits can lead to order one errors in the halo flattening and halo-to-disc force ratio at the Sun. We present a discussion of the dependence of these results on the progenitor mass, and demonstrate that the misalignment is mass-independent for the range of masses of observed streams. Hence, orbit-fitting does not yield better constraints on the potential if one uses narrower, lower-mass streams.
\end{abstract}


\section{Introduction}
The halo of the Milky Way is rich with substructure. Large optical surveys have revealed enhancements in the density of stars in the halo which trace out filaments across the sky. It is believed that each such filament is generated by stars being tidally stripped from a progenitor which has entered the influence of the Milky Way and hence these structures are called tidal streams. 

Tidal streams probe the outer parts of the Galactic potential, where the potential is expected to be dark-matter dominated. By understanding their formation it should be possible to constrain properties of the Galactic potential \citep{McGlynn1990,Johnston1996, Johnston1999}. One way of approaching this problem has been to assume the members of the stream delineate an orbit \citep{Jin2007,Binney2008}. If we are given phase-space coordinates for objects that lie at different phases of a single orbit, then the path of the orbit, and hence the underlying potential, may be recovered. Even if the observables are not well known, the orbit and underlying potential can be recovered with reasonable accuracy. If the data lie along an orbit then full six-dimensional phase-space information is redundant: \cite{EyreBinney2009A} showed that the orbit and potential could be recovered with positions, distances and line of sight velocities and \cite{EyreBinney2009B} did the same with proper motions instead of line of sight velocities. The technique of orbit fitting to stream data has been utilised most successfully by \cite{Koposov2010} who used data for the stream GD-1 \citep{GD1} to constrain a simple two-parameter logarithmic potential for the Milky Way as well as a more complex multi-component Milky Way potential.

The study of tidal streams has a natural expression in angle-action variables \citep{HelmiWhite1999,Tremaine1999,EyreBinney2011} and in the correct potential a stream should reveal a clear signature in angle-action space. \cite{EyreBinney2011} discussed and demonstrated the formation of streams in angle-action space through N-body simulations. The authors investigated the degree to which orbits delineate streams and found that assuming the stream lies along a single orbit can lead to systematic biases in estimates of the potential parameters. However, these authors were limited to quantifying the degree of misalignment for potentials in which the angle-action coordinates are analytically tractable (spherical and St\"ackel potentials). The effect of the misalignment in more realistic Galactic potentials remains an open question. In particular, is the misalignment for known streams in the Galaxy expected to significantly bias orbit-fitting algorithms?

In the next section we review the known tidal streams of the Milky Way and summarise the data from the literature which will be of use in answering this question. In Section~\ref{Formalism} we present the angle-action formalism and discuss the formation of tidal streams in this framework. In Section~\ref{Motivation} we motivate the need for an improvement on orbit-fitting algorithms by investigating the degree to which streams delineate orbits in realistic Galactic potentials, specifically focussing on which of the known streams can be reliably analysed using orbit-fitting algorithms. In Sections~\ref{MassDependence} and~\ref{Anisotropies} we discuss the validity of the presented formalism using N-body simulations and demonstrate the results obtained are independent of the progenitor's mass. We close by investigating the anticipated errors introduced by orbit-fitting when attempting to constrain the parameters of a two-parameter family of realistic Galactic potentials from stream data.

This paper is the first of two papers on the problem of stream-orbit misalignment. Motivated by the results of this paper, the second paper \citep[hereafter Paper II]{SandersBinney2013} presents an alternative to orbit-fitting based on the angle-action formalism.

\section{Known streams}\label{KnownStreams}
Before we discuss the theory of tidal streams and how they may be used to constrain the Galactic potential we give a short description of known long streams. It is important that we understand the available data before we concern ourselves with the details of analysing stream data. For each stream we have summarised the information from the literature that is useful for the following discussion. There are other streams, which we have not included. These streams are associated closely with globular clusters and dwarf galaxies and as such are short and not as useful for constraining the Galactic potential. The majority of the listed streams were discovered using matched-filter star counts \citep{Rockosi2002} on Sloan Digital Sky Survey data \citep[SDSS,][]{SDSS}.

\subsection{GD-1}
\cite{GD1} detected a $63^\circ$ stellar stream in SDSS data using star counts. This stream is referred to in the literature as GD-1. The stream is extremely narrow, from which the authors conclude that the progenitor was a globular cluster. However, the progenitor has not been identified suggesting that it has been completely disrupted. Because it is exceptionally long and thin, the GD-1 stream has been used by both \cite{Willett2009} and \cite{Koposov2010} to constrain the Galactic potential. Additionally, GD-1 is relatively close to the Sun for a tidal stream ($\sim 10\kpc$), which allowed these authors to construct a full 6D phase-space map of the stream. Both sets of authors used the assumption that the stream delineates an orbit. 

The data for GD-1 is currently the best data set for a tidal stream: \cite{Koposov2010} provides us with 6D phase-space coordinates for different fields along the stream. The authors fit an orbit to this stream using a 3-component potential. The best-fit orbit has its pericentre at $14\kpc$,  apocentre at $26\kpc$ and reaches a maximum height above the Galactic plane of $\sim 11\kpc$.

\subsection{Orphan}
The Orphan stream was discovered independently by both \cite{Grillmair2006-Orphan} and \cite{Belokurov2007-Orphan} using SDSS photometry and spectroscopy. The nearest part of the $50^\circ$-long stream is $\sim20\kpc$ from the Sun. The Orphan stream is so-called due to the lack of a progenitor. \cite{Belokurov2007-Orphan} suggested that Ursa Major II galaxy (UMa II) may be the progenitor. However, using distances and radial velocities \cite{Newberg2010} fitted an orbit to the stream which seemed to rule out UMa II as the progenitor. The more recently discovered star cluster Segue-1 \citep{Belokurov2007-Segue1} seems a more-likely candidate.

\cite{Newberg2010} find a best-fitting orbit for the Orphan stream with pericentre at $16.4\kpc$, apocentre at $90\kpc$ and reaching a maximum height above the Galactic plane of $\sim45\kpc$.

\subsection{Anticenter}
The Anticenter stream was detected by \citet{Grillmair2006-ASS}\ as a $\sim65^\circ$ long overdensity approximately $\sim9\kpc$ away in the direction of the Galactic anticentre. \cite{Grillmair2006-ASS} concluded that it was not associated with the Monoceros Ring, despite lying in the same region of the sky, and this conclusion was strengthened by the kinematics measured by \cite{Carlin2010}, who measured a 6D phase-space point of the stream. This single point may be used to construct an approximate orbit for the stream.

\subsection{NGC 5466}
A $45^\circ$ stream was detected by \citet{Grillmair2006-NGC5466}. It appears to coincide with the much smaller tidal tails of NGC 5466 found by \cite{Belokurov2005-NGC5466}, so it is believed to be associated with this extremely metal-poor globular cluster. In this case we are in the fortunate position of confidently identifying the progenitor and we may use the orbit of the progenitor as a proxy for the path of the stream. An approximate orbit for the progenitor may be constructed from the 6D coordinates of NGC 5466 given by \cite{Harris1996} and \cite{Dinescu1999}.

\subsection{Palomar 5}
Palomar 5 (Pal 5) is a very low mass, sparse halo cluster lying $18.6\kpc$ from the Sun, which was found to have short ($\sim2.5^\circ$), strong leading and trailing tidal tails by \cite{Odenkirchen2001}. It was the first example of tidal tails being resolved around a cluster and has received much attention in the literature as an example of the formation of tidal streams \citep{Dehnen2004}. Further observations found that the stream extended up to $22^\circ$ \citep{Grillmair2006-Pal5}. As with NGC 5466, we may use the 6D phase-space coordinates of the progenitor, given by \cite{Odenkirchen2001}, to construct an approximate orbit for the progenitor, and hence for the stream members.

\subsection{Sagittarius}
The Sagittarius dwarf galaxy was discovered by \cite{Ibata1995} and is the third largest satellite of the Milky Way. \cite{Johnston1995} predicted that the Sagittarius dwarf would be heavily disrupted, and that debris might be observed in the solar neighbourhood. \cite{Majewski2003} observed extended leading and trailing tidal tails, which \cite{Belokurov2006-FoS} found wrapped at least once around the Galaxy. Its length and number of constituent stars make the Sagittarius stream useful for constraining the Galactic potential. However, the Sagittarius stream is very broad and could potentially reflect the internal properties of the progenitor \citep{Penarrubia2010}. Complex models, which account for dynamical friction (see later), are required to use the Sagittarius stream to constrain the Galactic potential.

\cite{Belokurov2006-FoS} found that the Sagittarius stream had what they dubbed a bifurcation. These authors were limited to observing the stream in the northern Galactic hemisphere. Recently \cite{Koposov2012} have extended the observations to the southern Galactic hemisphere and found that the bifurcation is also present there. It is believed that the bifurcation is actually due to a fainter stream which runs alongside the Sagittarius stream. This secondary stream is chemically distinct from the Sagittarius stream \citep{Koposov2012}, which seems to rule out the possiblity that the secondary stream and the Sagittarius stream share a common progenitor. It is believed that the secondary stream originated from a different progenitor, presumably a companion of Sagittarius. 

We take the sky coordinates of the Sagittarius dwarf galaxy from \cite{Majewski2003}, the distance from \cite{Siegel2007}, the line-of-sight velocity from \cite{Ibata1997} and the proper motions from \cite{Pryor2010}, giving us a 6D phase-space point on the orbit of Sagittarius.
 
\subsection{Acheron, Cocytos, Lethe and Styx}
Four streams were discovered by \cite{Grillmair2009} using a matched-filter technique and were named Acheron, Cocytos, Lethe and Styx in order of increasing distance from the Sun. The first three of these are very narrow and lie between $3$ and $15\kpc$ from the Sun spanning between $37^\circ$ and $84^\circ$. Styx is much more distant ($\sim 45\kpc$), broader and spans at least $53^\circ$. None of the four streams has an identified progenitor although the Styx stream is believed to be associated with the concurrently discovered cluster Bootes III. In the discovery paper \citeauthor{Grillmair2009} fits orbits to the available data to predict a 6D phase-space point in each of the streams.

\subsection{Aquarius}
The Aquarius stream was detected as an overdensity in the line-of-sight velocity data from the Radial Velocity Experiment (RAVE) by \cite{Williams2011}. The stream passes very close to the Sun (within $0.5\kpc$), is particularly broad and has no identified progenitor. \cite{Williams2011} fits an orbit to this stream with pericentre at $1.8\kpc$,  apocentre at $9.0\kpc$ and reaching a maximum height above the Galactic plane of $\sim5\kpc$.

\subsection{Cetus, Virgo and Triangulum}
The Cetus stream was discovered by \cite{Newberg2009} in velocities from the Sloan Extension for Galactic Understanding and Exploration \citep[SEGUE,][]{Yanny2009}. These observations were corroborated by \cite{Koposov2012} who observed the Cetus stream in the SDSS southern Galactic hemisphere data. The stream lies $\sim34\kpc$ from the Sun and follows an approximately polar orbit. 

\cite{Juric2008} discovered a faint overdensity in the constellation of Virgo from SDSS stellar number counts. The Virgo overdensity was also observed as a velocity overdensity in measurements of RR Lyrae stars \citep{Duffau2005}. The overdensity has a large spatial extent but it is unclear whether it is a stream or not. 

The Triangulum stream was very recently discovered by \cite{Bonaca2012} by searching SDSS data using a matched-filter technique. The stream extends over $12^\circ$ and lies approximately $26\kpc$ from the Sun.

These three streams do not have sufficient data in the literature to reliably construct their 6D phase-space structure. A simple orbit fit may be possible but this is beyond the scope of this exercise.

\section{Tidal streams in angle-action coordinates}\label{Formalism}
\cite{Tremaine1999} and \cite{HelmiWhite1999} explained the formation of tidal streams in angle-action space. In this formulation streams are formed because stars do not share a common orbit. It is this formulation which we present here.

Given 6D phase-space information, the angle-action coordinates for each star along the tidal stream may be found. We assume that each star does not feel the gravitational influence of the stars in the stream but only the external Galactic potential. Angle-action variables provide a simple way to follow the dynamics of the stream as the actions are constants of the motion whilst the angles increase linearly in time. For a single star in the stream, the angle-action coordinates, $(\boldsymbol{\theta},\boldsymbol{J})$, obey the equations
\begin{equation}
\boldsymbol{J}=\text{const.},\quad\boldsymbol{\theta}(t)=\boldsymbol{\theta}(0)+\boldsymbol{\Omega}t,
\end{equation}
where $\boldsymbol{\Omega} = \partial H/\partial \boldsymbol{J}$ are the frequencies of the Hamiltonian, $H$, and $t$ is the time since the star was stripped from the progenitor. All the stars in the stream are assumed to derive from a progenitor with actions $\boldsymbol{J}_0$. The progenitor is assumed to be of low mass so that we may neglect dynamical friction, and the actions of the progenitor are constant throughout the motion. Also, we assume that once a star has been stripped, the influence of the progenitor can be neglected so the star's actions are constant from the time the star was stripped.

The stream is formed by the difference in angles between the progenitor and the stars in the stream, $\Delta\boldsymbol{\theta}$, increasing with time. For a single star we have
\begin{equation}
\Delta \boldsymbol{\theta} = \boldsymbol{\theta}-\boldsymbol{\theta}_0 = \Delta \boldsymbol{\Omega}t + \Delta\boldsymbol{\theta}(0).
\end{equation}
$\Delta\boldsymbol{\theta}(0)$ is the initial difference in angles between the progenitor and a given star. $\Delta\boldsymbol{\Omega}$ is the difference in frequencies. Both $\Delta\boldsymbol{\theta}(0)$ and $\Delta\boldsymbol{\Omega}$ depend upon the progenitor mass (see Section~\ref{MassDependence}). When a stream has formed, $\Delta\boldsymbol{\theta}(0)$ has become small compared to the term $\propto t$ so we have
\begin{equation}
\Delta \boldsymbol{\theta} \approx\Delta \boldsymbol{\Omega}t.
\label{Theta-FreqCorrelation}
\end{equation}
As the frequencies and the angles depend on the potential, this equation provides a constraint for the potential. However checking whether this equation is obeyed for all stars in the stream in a given potential is complicated for two reasons:
\begin{inparaenum}
\item The progenitor of the stream may be unknown, and
\item the time that the star left the progenitor, $t$, is not known.
\end{inparaenum}


Nevertheless, equation~\eqref{Theta-FreqCorrelation} provides a useful constraint on the potential. If the difference between the actions of the stream stars and those of the progenitor is small, the frequencies of a stream star are well approximated by the Taylor expansion
\begin{equation}
\boldsymbol{\Omega} \approx \boldsymbol{\Omega}_0 + \boldsymbol{D}\cdot\Delta\boldsymbol{J},
\label{OmegaTS}
\end{equation}
where $\boldsymbol{D}$ is the Hessian matrix
\begin{equation}
D_{ij}(\boldsymbol{J})=\frac{\partial^2 H}{\partial J_i \partial J_j}.
\end{equation}
This matrix is symmetric, so at each point of action space it is characterised by three orthogonal eigenvectors, $\hat{\boldsymbol{e}}_i$, with associated real eigenvalues, $\lambda_i$. With the Taylor series for the frequencies given in equation~\eqref{OmegaTS}, the difference in angles is related to the difference in actions by
\begin{equation}
\Delta\boldsymbol{\theta}\approx\Delta\boldsymbol{\Omega} t\approx\boldsymbol{D}\cdot\Delta\boldsymbol{J} t.
\end{equation}
In this framework we can understand the conditions required for a stream to form. Once a star has been stripped from the cluster, the action difference, $\Delta\boldsymbol{J}$, is frozen in and the angle difference increases with time. The Hessian determines along which directions the cluster spreads. For a long thin stream to form from an approximately isotropic cluster in action space, one eigenvalue of the Hessian must be much larger than the other two, $\lambda_1\gg\lambda_2\geq\lambda_3$. In this case the stream will stretch along the eigenvector $\hat{\boldsymbol{e}}_1$ and
\begin{equation}
\frac{\Delta\boldsymbol{\theta}}{t}\approx\Delta\boldsymbol{\Omega}\approx \hat{\boldsymbol{e}}_1 (\lambda_1 \hat{\boldsymbol{e}}_1\cdot\Delta\boldsymbol{J}).
\label{AngFreqAlignment}
\end{equation}
Hence the frequency difference should be aligned with the principal eigenvector of the Hessian for all stars in the stream, independent of their action. 

\section{The problem with orbit-fitting}\label{Motivation}
We have seen that if the Hessian matrix is dominated by a single eigenvalue, $\lambda_1$, the stream will stretch along the corresponding eigenvector, $\hat{\boldsymbol{e}}_1$. In general this vector will be misaligned with the progenitor frequency vector, $\boldsymbol{\Omega}_0$, with the angle, $\varphi$, between them given by
\begin{equation}
\varphi \equiv \arccos\Big(\hat{\boldsymbol{\Omega}}_0\cdot\hat{\boldsymbol{e}}_1\Big).
\end{equation}
The misalignment between these two vectors gives an indication of the error expected when the Galactic potential is constrained by assuming that the stream delineates an orbit. The potential in which the stream appears to delineate an orbit will, in general, be different to the true potential. Orbit-fitting algorithms also assume that the actions of all the constituent stars are the same which can also lead to errors. However, this effect is small, as the stream spans a small range in actions, so we assume that the misalignment angle gives rise to all the error in orbit-fitting algorithms. Importantly, this misalignment is independent of the progenitor mass, and depends only on the progenitor orbit, and hence the underlying potential (see Section~\ref{MassDependence}). Moving to lower-mass, and hence narrower streams, does not decrease the misalignment.

One key result of \cite{EyreBinney2011} is that when this misalignment is $\varphi=1.5^\circ$ in the isochrone potential, the mass of the Galaxy is overestimated by approximately $20\percent$ when using an orbit-fitting algorithm. Thus, even a small value of $\varphi$ can lead to significant error in the potential. \cite{EyreBinney2011} found the misalignment angle to be $\varphi\approx1-3^\circ$ at every point in action space in the isochrone potential, whereas in a St\"ackel potential the angle was as large as $\varphi\approx20^\circ$. These are special cases so it is necessary to explore the magnitude of this angle for realistic Galaxy potentials to assess the need to go beyond orbit-fitting algorithms.

We calculate $\varphi$ by first finding the Hessian matrix at each point in action-space. This calculation is simply performed using the torus machine \citep{McMillanBinney2008}. The torus machine constructs orbital tori of given actions for a general potential. Position and velocity coordinates are determined as functions of the angles on the surface of the torus. Given a set of actions the torus machine returns the corresponding frequencies. It is these properties which make it an appropriate tool for this task. 

For each point in action-space we use the torus machine to differentiate estimates of the frequency numerically. The error in the estimated actions of points that lie on a torus created by the torus machine is estimated as
\begin{equation}
\Delta J \approx \frac{\Delta H}{\sqrt{\Omega_R^2+\Omega_z^2}}
\end{equation}
where $\Delta H$ is the rms variation in the energy across the torus and $\Omega_i$ are the frequencies. For each action-space point we create a torus with the required actions, $\boldsymbol{J}$, and accuracy, $\Delta J$, as well as neighbouring tori which lie $\delta J$ away from the action-space point in each action-space direction. We use the calculated frequencies of these tori to construct numerically the Hessian matrix, $\partial\Omega_i/\partial J_j$, at the action-space point. We require $\Delta J\ll\delta J$ to ensure the numerical differentiation is accurate. The angle between the principal eigenvector of this matrix and the frequency vector at the action-space point is $\varphi$. We also calculate $\lambda_1/\lambda_2$ which gives a measure of the width of a stream formed at this action-space point. If this ratio is large, a long thin stream forms and equation~\eqref{AngFreqAlignment} is satisfied. However, if the ratio is small, the stream will be broad.

For a given choice of $\Delta J$ and $\delta J$ the error in $\varphi$ is estimated by calculating it for a known case. In the Kepler potential the angle is zero \citep{EyreBinney2011} and we use this fact to estimate the error in $\varphi$ in a general potential.

\begin{figure*}
$$\includegraphics{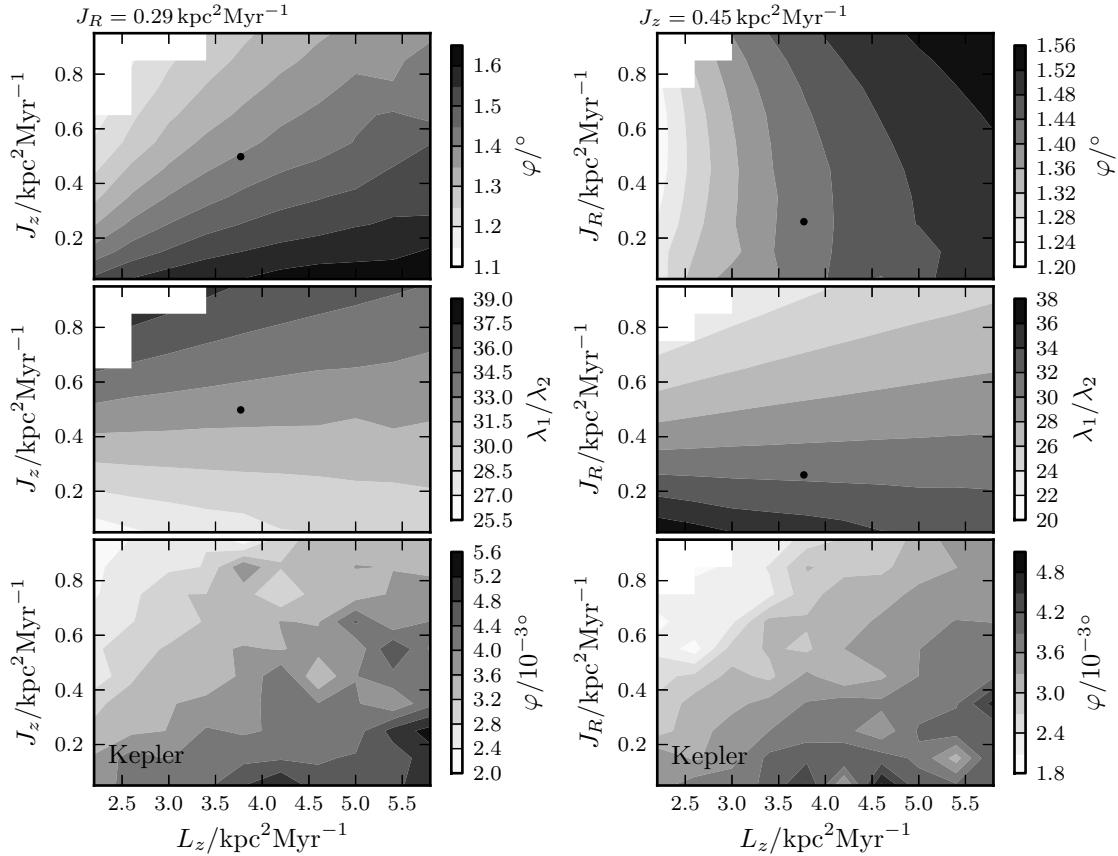}$$
\caption{Misalignment angle and ratio of two largest eigenvalues for the logarithmic potential with $V_c=220\kms$ and $q=0.9$. Two planes are displayed: $J_R=0.26\kpcMyr$ and $J_z=0.45\kpcMyr$. The bottom two panels give the misalignment angle in the Kepler potential. This should be zero everywhere so gives a measure of the error in the misalignment angle calculated in the logarithmic potential. The black dot shows the approximate action coordinates of GD-1, which the simulations in Section~\ref{MassDependence} and Paper II were chosen to emulate.}
\label{Log}
\end{figure*}
\begin{figure*}
$$\includegraphics{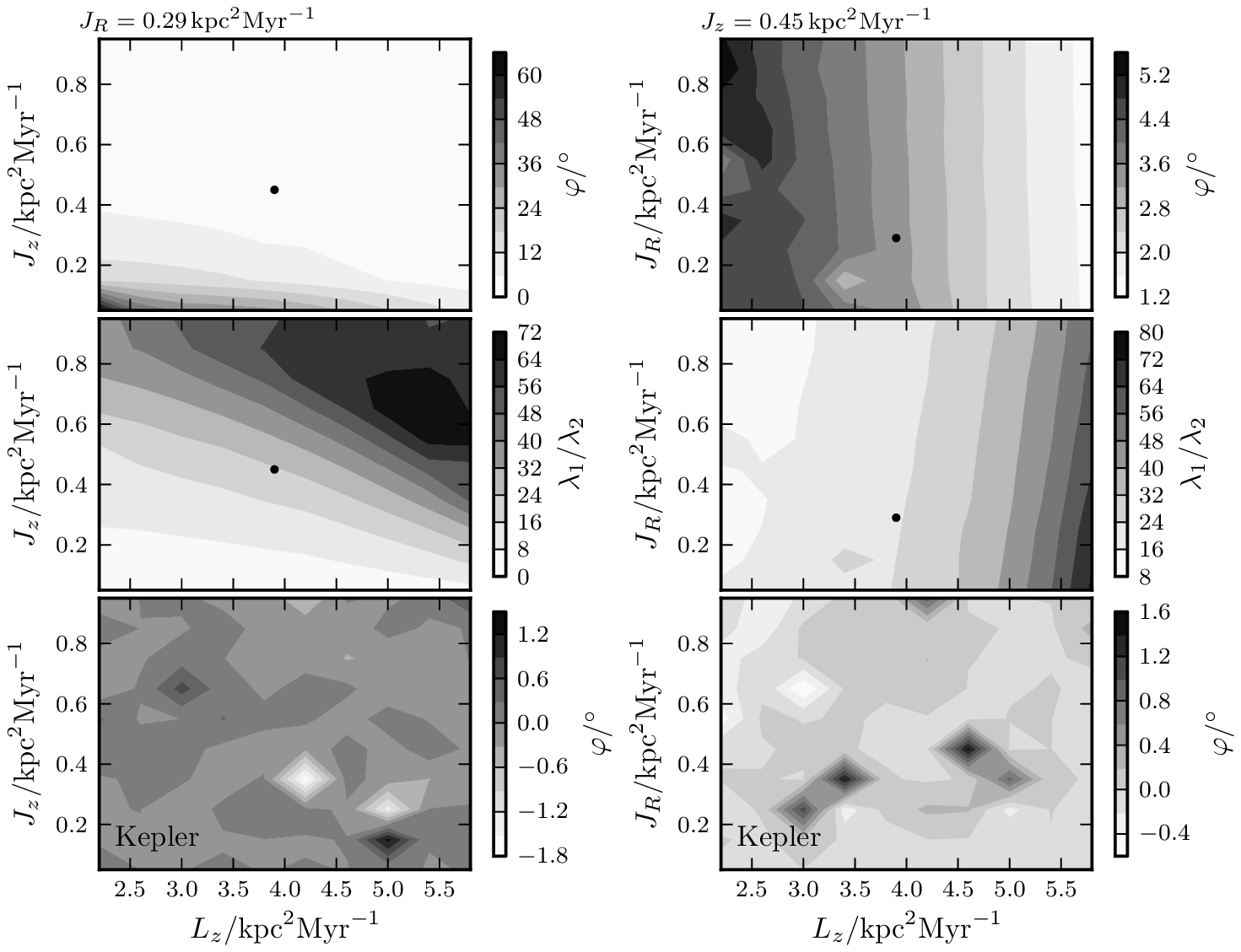}$$
\caption{Misalignment angle and ratio of two largest eigenvalues for McMillan's best potential (PJM11). Two planes are displayed: $J_R=0.25\kpcMyr$ and $J_z=0.15\kpcMyr$. The bottom two panels give the misalignment angle in the Kepler potential. This should be zero everywhere so gives a measure of the error in the misalignment angle calculated in the PJM11 potential. The black dot shows the approximate action coordinates of GD-1, which the simulations in Section~\ref{MassDependence} and Paper II were chosen to emulate.}
\label{PJM11}
\end{figure*}

We perform the above procedure on two potentials: the two-parameter logarithmic potential and the best-fit potential from \cite{McMillan2011} (hereafter referred to as PJM11). The logarithmic potential is defined as
\begin{equation}
\Phi(R,z) = \frac{V_c^2}{2}\ln\Big(R^2+\frac{z^2}{q^2}\Big),
\label{LogPot}
\end{equation}
where $V_c$ is the asymptotic circular speed and $q$ is the flattening parameter. We choose $V_c=220\kms$ and $q=0.9$ which gives a good representation of the potential of the Milky Way \citep{Koposov2010}.
The PJM11 potential is a multi-component potential generated by a bulge, thick and thin discs and a halo, which has been fitted to current experimental constraints. Fig.~\ref{Log} and Fig.~\ref{PJM11} show $\varphi$ and the eigenvalue ratio in two action planes for these two potentials\footnote{Throughout this paper the actions are stated in units of $\ActionUnits = 977.8\kpc\kms$}.


For the logarithmic potential $\varphi$ is small (about $1.5^\circ$), but crucially non-zero at all points in the action-space planes explored. This is similar to the value found by \cite{EyreBinney2011} for the isochrone potential. 
$\varphi$ decreases with increasing $J_z$ and increasing $L_z$ as orbits move further out in the potential, but the trend is very subtle. The errors in $\varphi$ are $\sim 0.005^\circ$. The eigenvalue ratio, $\lambda_1/\lambda_2$, is greater than $20$ for all action-space points shown. Therefore, we expect long thin streams which are misaligned with the orbit of their progenitor.

For the PJM11 potential $\varphi$ is $\ga 1^\circ$ for all the action-space points shown in Fig.~\ref{PJM11}, and can be as large as $\sim 40^\circ$ for orbits with low $J_z$. These low-$J_z$ orbits are planar disc-like orbits so clearly the discs in the PJM11 potential have a large effect on the misalignment angle. For these orbits the ratio of eigenvalues is small so broad structures will form from debris stripped from disc-like orbits. $\varphi$ decreases rapidly with increasing $J_z$ such that for orbits which spend most of their time out in the halo, $\varphi$ has a similar value to that found for the logarithmic and isochrone potentials. Similarly the ratio of the eigenvalues increases with increasing $J_z$. In the halo, long thin streams will form. The planes of constant $J_z$ show that orbits with large $L_z$ will form narrower streams which are more aligned with their progenitor orbit. Interestingly, despite the magnitudes of $\varphi$ being similar far out in the halo, the shapes of the $\varphi$ surface and the $\lambda_1/\lambda_2$ surface are very different for the two potentials.

In both potentials examined here, $\varphi$ is a few degrees far out in the halo. However, for the PJM11 potential, this can increase to several tens of degrees for orbits which spend more time in the disc. Therefore, simulations which use the logarithmic potential may not give a good representation of the evolution of a tidal stream. The stream will potentially delineate the orbit more than it should in a realistic potential giving the impression that orbit-fitting algorithms are appropriate. We note that both the logarithmic potential used here and the PJM11 potential have approximately spherical halo potentials. With more halo flattening we expect the misalignment angle will increase \citep[c.f. the St\"ackel potential analysed by][]{EyreBinney2011}. 

\subsection{Known streams}
We have seen that $\varphi$ is definitely non-zero and can be large for realistic Galaxy potentials, and thus systematic errors can be made when an orbit-fitting algorithm is used. We now explore the magnitude of the misalignment angle for the known streams described in Section~\ref{KnownStreams} to decide whether orbit-fitting algorithms are appropriate for analysing available stream data. For this task we must make estimates of their actions and then use the torus machinery as above.

With the information collected in Section~\ref{KnownStreams} we can construct approximate 6D phase-space points for each of the streams. For those streams with known progenitors (Palomar 5, Sagittarius, NGC 5466) we use the 6D coordinates of the progenitor. The Acheron, Cocytos, Lethe and Styx streams have predicted 6D stream points from an orbit fit by \cite{Grillmair2009}. The Anticenter stream has a single measured 6D stream coordinate. The remaining streams (GD-1, Orphan, Aquarius) have approximate orbit fits from the literature. In our chosen potential we can produce similar orbits and find a single 6D point on these orbits.

The actions in the PJM11 potential for each of these points are found using the St\"ackel-fitting algorithm from \cite{Sanders2012} (this algorithm is briefly discussed in the appendix of Paper II). There is an error of $\sim 5-10\percent$ in the actions introduced by the St\"ackel-fitting algorithm, but this is irrelevant when compared with the observational uncertainties in the coordinates of these streams. With these actions we find $\varphi$ to a precision of better than $0.1^\circ$ using the torus machine. For each stream Table~\ref{StreamActionTable} gives the approximate actions, the ratio of the two largest eigenvalues of the Hessian and $\varphi$. The misalignment angle varies from $\varphi\approx 13^\circ$ to $\varphi\approx0.15^\circ$ with the largest misalignment angle producing the smallest eigenvalue ratio, and hence the broadest streams. 

The Anticenter, Aquarius and NGC 5466 streams all have $\varphi>2.8^\circ$ and small eigenvalue ratios ($\sim 6$). Therefore, just from their actions we anticipate that the streams formed will be broad. This is definitely true of the Aquarius stream, and the Anticenter stream is a complex which is believed to consist of three separate streams with the whole complex having a width of $\sim5^\circ$ \citep{Grillmair2006-ASS}. For all other streams we find $\varphi\la1^\circ$ and the eigenvalue ratio is large ($>20$) so narrow streams are expected. 

\section{Mass Dependence}\label{MassDependence}
When presenting the angle-action formalism of stream formation we made little mention of the progenitor mass. We would like to know for what range of progenitor masses this approach and the above results are valid. Stream progenitor masses span a large range: GD-1 is observed to have a mass of $~2\times10^4M_\odot$ \citep{Koposov2010}, whilst the Sagittarius dwarf is believed to be $~10^8-10^9M_\odot$ \citep{Law2005,Fellhauer2006}.

Here we discuss each of the assumptions made in the angle-action formalism in the context of mass-dependence:
\begin{inparaenum}
\item The progenitor actions are assumed to be constant for all time. This is valid in the limit that dynamical friction is negligible. We can neglect dynamical friction if we are in the regime where
\begin{equation}
M_c\ll\frac{r_p V_c^2}{G},
\end{equation}
where $V_c$ is the circular speed of the potential, and $r_p$ is the pericentre radius. This mass is approximately $10^{11}M_\odot$ for a GD-1-like orbit, so we expect this effect to be negligible for $M_c\lesssim 10^9M_\odot$. Many streams lie much further out in the halo where this limit is expected to be much larger. Additionally, the effects of dynamical friction are expected to be comparable for the cluster and the stream, such that the relative structure of the stream is not affected, but the global cluster-stream complex is. We expect that dynamical friction is irrelevant for most streams, but its effects on the Sagittarius stream may be important \citep{JiangBinney2000}.
\item We have assumed that a particle is instantaneously released from the cluster, and subsequently has constant actions. However, the self-gravity of the cluster will always be significant, regardless of mass. It is the self-gravity of the cluster which determines whether a particle leaves the cluster on each pericentric passage. A particle will always leave the cluster in the same way (approximately through the Lagrange points at pericentre), irrespective of the mass. In the absence of self-gravity we do not have this restriction as particles leave the cluster more uniformly. Therefore, we expect that the inclusion of self-gravity will have an impact on the overall shape of the angle-action space structure of the stream, independent of the mass (see~\ref{Anisotropies}). We also expect that increased cluster self-gravity will produce broader streams as particles leave the cluster with a larger range of actions. Thus the role of progenitor mass is to set the scale of the stream's structure without affecting its morphology in any other way. Despite the cluster self-gravity always being important, it should produce mass-independent effects on the overall shape of the stream (see below).
\item We assume that we can neglect the finite angle size of the cluster, $\Delta\boldsymbol{\theta}(0)$ after some time $t$ as it is negligible compared to the contribution of secular evolution to $\Delta\boldsymbol{\theta}$. Assuming the secular evolution of the stream stretches the angle distribution along one direction (see below) $\Delta\boldsymbol{\theta}(0)$ will act to broaden the stream perpendicular to this principal direction. This is mass-dependent, and is related to the above self-gravity arguments, but when a stream has formed this term is always unimportant. As long as initial spread in angles is symmetric about the stream path this term will not affect the presented formalism.
\item We employ a Taylor expansion in $\Delta\boldsymbol{J}$ when finding the relationship between the frequencies and actions. This is important as it leads to the conclusion that $\Delta\boldsymbol{\Omega}$ for each particle will lie along the same vector $\hat{\boldsymbol{e}}_1$. This assumption is valid provided 
\begin{equation}
D_{ij} \gg \frac{\partial D_{ij}}{\partial J_k}\Delta J_k.
\label{TaylorValidity}
\end{equation}
If the Hamiltonian is a function of some low power of $\boldsymbol{J}$ then this reduces to 
\begin{equation}
\Delta J_i \ll J_i.
\end{equation}
\end{inparaenum}

If we are in the progenitor-mass regime where we can neglect the above effects, what effect does the progenitor mass have on the resulting angle-action space distribution? $\Delta\boldsymbol{J}$, $\Delta\boldsymbol{\Omega}$ and $\Delta\boldsymbol{\theta}$ are all functions of the progenitor mass. We expect that larger progenitor masses produce larger spreads in the actions, frequencies and angles of the resulting stream, but we would like to know their exact mass dependence. Following \citet{EyreBinney2011} we have that 
\begin{equation}
\Delta J_i \approx \frac{1}{2\pi}\oint \Delta p_i \mathrm{d}x_i \approx \frac{1}{2\pi}\oint \sigma \mathrm{d}x_i,
\end{equation}
where $\sigma$ is the velocity dispersion of the progenitor. For axisymmetric systems this is approximately
\begin{equation}
\boldsymbol{\Delta J} \approx \frac{1}{2\pi} (2 \sigma\Delta R, 2\pi \sigma r_p, 4\sigma\Delta z)
\end{equation}
where $\Delta R$ is the difference between the apocentric and pericentre radius, and $\Delta z$ is the maximum height above the plane reached by the orbit. $\Delta L_z$ is calculated at pericentre as this is where the majority of particles are stripped. Under the assumption of Equation~\eqref{TaylorValidity}, $\Delta\boldsymbol{\Omega}$ is linearly related to $\Delta\boldsymbol{J}$ via the (mass-independent) Hessian $D_{ij}$, and $\Delta\boldsymbol{\theta}$ is linearly related to $\Delta\boldsymbol{\Omega}$ via the time since stripping, $t$. Therefore, both $\Delta\boldsymbol{\Omega}$ and $\Delta\boldsymbol{\theta}$ will have the same dependence on mass as $\Delta\boldsymbol{J}$. 

From the virial theorem we relate the velocity dispersion of the cluster to its mass and radius via
\begin{equation}
\sigma^2 \approx \frac{GM_c}{r_c}.
\end{equation}
and the tidal radius, $r_c$, is related to the mass of the cluster via 
\begin{equation}
r_c=r_p\Big(\frac{M_c}{M_g}\Big)^\frac{1}{3},
\label{TruncationRadius}
\end{equation}
where $M_g$ is the mass of the host galaxy contained within $r_p$. Therefore, the progenitor mass is proportional to the velocity dispersion cubed or $\sigma\propto M_c^{1/3}$. As $\Delta\boldsymbol{J}$, $\Delta\boldsymbol{\theta}$ and $\Delta\boldsymbol{\Omega}$ are proportional to $\sigma$ in the regime we are considering, we expect all these quantities to also depend on $M_c^{1/3}$. \cite{Choi2007} showed from N-body simulations that the energy difference of stripped particles obeyed this same scaling with progenitor mass in a spherical halo. Similarly, \cite{Johnston1998} demonstrated that the density profile along a stream was described by the same analytic form scaled by $M_c^{1/3}$, and \cite{Johnston2001} utilised this scaling relation to develop a semi-analytic formalism for predicting the morphology of a recently formed stream.

These arguments convince us that the progenitor mass acts only to scale the angle-frequency distribution, and the shape is independent of the mass, provided we are in the aforementioned regime. Therefore, the misalignment angle is mass-independent.

Using these results, we relate the assumption of Equation~\eqref{TaylorValidity} to a constraint on the progenitor mass for a given orbit. We expect the neglected terms in the Taylor series to be non-negligible when
\begin{equation}
\Delta\boldsymbol{J} \approx \frac{1}{2\pi} (2 \sigma\Delta R, 2\pi\sigma r_p , 4\sigma\Delta z) = \boldsymbol{J}.
\end{equation}
Therefore, we expect the assumption to break down when
\begin{equation}
\sigma\gtrsim{\rm min}\{\frac{\pi J_R}{\Delta R}, \frac{L_z}{r_p}, \frac{\pi J_z}{2\Delta z}\},
\label{SigMax}
\end{equation}
where all the quantities on the right-hand side depend only on the chosen orbit. We see that the cluster needs to be on a sufficiently eccentric orbit for the approximation to hold. However, we expect that the majority of tidal streams are formed from progenitors on eccentric orbits, so this constraint is not too restrictive.

All the above predictions may be tested by inspecting some N-body simulations. We construct a stream by placing a King cluster at apocentre on a stream-like orbit in the logarithmic potential defined by equation~\eqref{LogPot} and integrating with self-gravity until a stream has formed. King models \citep{King1966} are characterised by three free parameters: the ratio of central potential to squared-velocity parameter, $W_0 = \Psi_0/\sigma^2 = 2$, the cluster mass, $M_c$, and a tidal limiting radius, $r_t$, set by equation~\eqref{TruncationRadius}. We seed the clusters with $N=10000$ particles, and explore the range of masses $2\times10^4\leq M_c \leq 2\times10^9 M_\odot$. The parameters for the simulations are given in Table~\ref{KingModels}.
\begin{table}
\caption{Parameters of King models used in the simulations detailed in Section~\ref{MassDependence}. $\epsilon$ is the softening parameter.}
\begin{tabular}{lllllll}
$N$&$W_0$&$r_p/\kpc$&$\frac{M_c}{M_\odot}$&$r_t/\kpc$&$\sigma/\kms$&$\epsilon/\pc$\\
\hline\\
$10000$&$2.0$&$14$&$2\times10^4$&$0.07$&$1.39$&$1.5$\\
&&&$2\times10^5$&$0.14$&$3.01$&$3$\\
&&&$2\times10^6$&$0.32$&$6.50$&$6$\\
&&&$2\times10^7$&$0.69$&$14.0$&$14$\\
&&&$2\times10^8$&$1.48$&$30.1$&$30$\\
&&&$2\times10^9$&$3.20$&$65.0$&$66$\\
\hline
\end{tabular}
\label{KingModels}
\end{table}

The orbit was chosen to be similar to the orbit of the GD-1 stream \citep{Koposov2010}. The orbit has initial conditions $(R,z)=(26.0,0.0)\kpc$ and $(U,V,W)=(0.0,141.8,83.1)\kms$ where positive $U$ is towards the Galactic centre and positive $V$ is in the direction of the Galactic rotation at the Sun. This orbit has $r_p \approx 14\kpc$. We evolve the simulation for $t=4.27\Gyr$ (just after the $11$th pericentric passage) using the code {\textsc{gyrfalcON}} \citep{Dehnen2000,Dehnen2002}, made available through the NEMO Stellar Dynamics Toolbox \citep{NEMO}.

\begin{figure}
$$\includegraphics{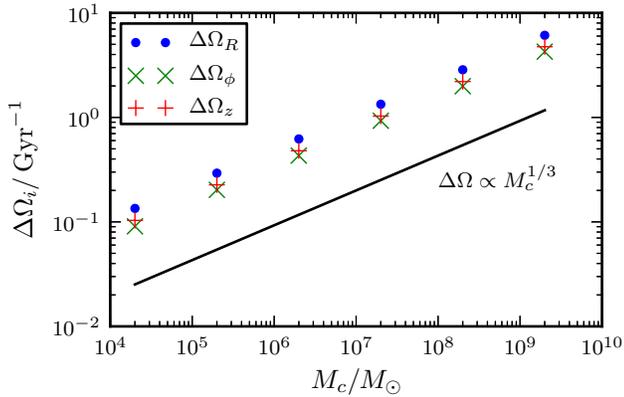}$$
\caption{The size of the frequency distribution against the progenitor mass. The size is estimated using the standard deviation in each frequency coordinate. As expected, $\Delta\Omega_i$ is proportional to $M_c^{1/3}$.}
\label{DeltaFreq}
\end{figure}

\begin{figure}
$$\includegraphics{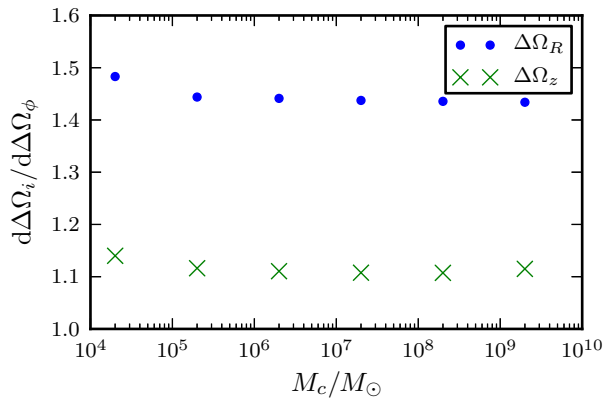}$$
\caption{The slope of the frequency distribution against the progenitor mass. We see the slopes are independent of mass. There is a small deviation at low mass due to errors introduced by the St\"ackel-fitting algorithm used to estimate the frequencies. At the high mass end there is also a slight deviation which may be due to the higher-order action space structure.}
\label{Gradient}
\end{figure}

For each resulting particle distribution we cut out the remnant of the progenitor, and estimate the actions, angles and frequencies of the stream particles using the St\"ackel-fitting algorithm \bibpunct[; ]{(}{)}{;}{a}{}{;}\citep[Paper II]{Sanders2012}\bibpunct[; ]{(}{)}{;}{a}{}{,}. We quantify the spread in each coordinate using the standard deviation. In Fig.~\ref{DeltaFreq} we plot the frequency difference as a function of progenitor mass. The correlation is very tight and, from the guiding line with slope $1/3$ we see the data follow the expected trend. In Fig.~\ref{Gradient} we plot the gradient of the frequency distribution. It is this quantity which gives the degree of stream-orbit misalignment. We see that, as expected, the gradient is constant with mass. There is a small deviation at the low mass end, which is due to the numerical errors introduced by the St\"ackel-fitting algorithm. We have near-perfect scaling of the results with mass, so there are no mass-dependent effects in the mass regime considered.

For this orbit we use equation~\eqref{SigMax} to find the maximum progenitor velocity dispersion for the first order expansion of $\Delta\boldsymbol{\Omega}$ in terms of $\Delta\boldsymbol{J}$ to be valid. We find that radial and vertical actions give similar constraints of $\sigma_{\rm max}\approx70\kms$ which translates into a maximum mass of $M_{\rm max}\approx5\times10^9M_\odot$. We have not quite reached this regime with the N-body simulations, but there is the suggestion of its impact at the high-mass-end of Fig.~\ref{Gradient}. 

We have seen that the formalism is valid for $M_c\lesssim10^{9}-10^{10}M_\odot$, when dynamical friction, the higher order action-space structure and perturbations from the cluster remnant become important. However, below this limit we find that the progenitor mass acts to scale the frequency, action and angle-distributions, such that the shapes of these distributions are essentially mass-independent. Therefore, we expect the angle-action formalism and the results of the previous sections to be valid for all observed streams, although dynamical friction may be relevant for the Sagittarius stream.

\section{Anisotropies in the action distribution}\label{Anisotropies}

In Section~\ref{Motivation} we showed that $\varphi$ is non-zero for the logarithmic potential. From the above simulations $\varphi=0.18^\circ$ but using the torus machine we find that $\varphi=1.83^\circ$. The source of this discrepancy is found by running the simulation without self-gravity. In that case $\varphi=1.92^\circ$ in agreement with the prediction. Therefore, the self-gravity of a cluster causes $\varphi$ to decrease. The gradient of the frequency distribution is mass-independent, so this self-gravity effect is also mass-independent.

\begin{figure}
$$\includegraphics{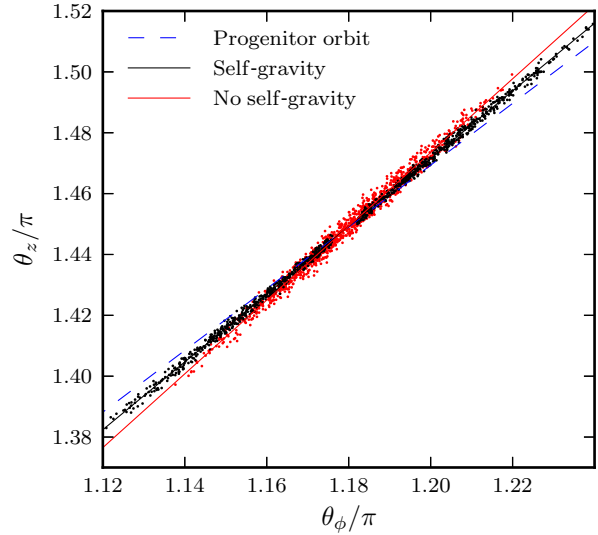}$$
\caption{Angle-angle plot for a similar simulation to those outlined in Section~\ref{MassDependence} but in the PJM11 potential run with and without self-gravity. This snapshot shows the clusters at the fifth apocentric passage. The red line gives the orbit of the progenitor. The black and blue lines are straight line fits to the data for the self-gravity and no self-gravity simulations respectively. We note that there is a misalignment between the stream and the progenitor orbit which decreases when self-gravity is included. The same effect is observed in the logarithmic potential, but it is clearer to see in the PJM11 potential.}
\label{GravNoGrav}
\end{figure}

A similar experiment run in the PJM11 potential shows a similar $\sim1.5^\circ$ decrease in $\varphi$ (see Fig.~\ref{GravNoGrav}). However, in this case, the simulation with gravity included still shows a significant $\varphi$. It just seems a coincidence that for the simulation in the logarithmic potential the expected value of $\varphi$ is almost cancelled by the inclusion of self-gravity.

This effect can be understood by considering the action-space structure of the cluster \citep{EyreBinney2011}. In the formalism of Section~\ref{Formalism} we showed that the stream would lie along the pirncipal eigenvector of the Hessian, but only if the stream action-space distribution is isotropic. \citet{EyreBinney2011} showed that the action-space distribution is not isotropic. Self-gravity introduces different anisotropies to those present when self-gravity is neglected. Different action-space distributions will give rise to different frequency-space distributions under the action of the Hessian. For Hessians with large eigenvalue ratios a highly elongated frequency distribution will be produced, but its orientation will depend on the shape of the action-space distribution.

We can understand the difference between the self-gravity and no-self-gravity simulations by considering how the particles are stripped from the cluster when self-gravity is included. For a particle to be stripped it must leave the cluster through the Lagrange points L1 and L2 at pericentre. For the orbit considered, the motion is dominated by the radial motion, and at pericentre the Lagrange points lie in a plane which is nearly parallel to the plane $z=0$. Therefore, a particle needs to have an increase in its radial velocity to have sufficient energy to be stripped. This increased radial motion will in turn increase/decrease the radial frequency $\Omega_R$ depending on whether the particle leaves through L1 or L2. Similarly the angular frequency $\Omega_\phi$ will increase/decrease as the particle moves to a smaller/larger radius without changing its transverse velocity. If we consider the motion in $z$ to be completely decoupled from the radial motion, which in the orbit considered is a fair assumption, increased motion in $R$ as the particle leaves the cluster will not alter the vertical action $J_z$ and frequency $\Omega_z$.


Now we can understand Fig.~\ref{GravNoGrav} as the result of this frequency-space evolution. Particles in the stream have increased/decreased angular frequency whilst their vertical frequency has remained constant. This causes the distribution in $(\theta_\phi,\theta_z)$ space to rotate clockwise thus decreasing the misalignment angle.

We investigate how the anisotropy of the action-space distribution affects the estimated misalignment angle for the presented known streams. We could attempt to estimate the effects of the anisotropy analytically following a similar analysis to \citet{EyreBinney2011}. However, as we are only dealing with eleven streams, we choose to run some N-body simulations, which will fully account for these effects. For each known stream we integrate the orbit in the PJM11 potential to find the pericentre radius, $r_p$, and a phase-space point at apocentre. We seed a $10000$ particle $2\times10^5 M_\odot$ King cluster with a tidal radius related to $r_p$ via equation~\eqref{TruncationRadius}, and place it at the apocentre phase-space point. The simulation is then evolved in {\textsc{gyrfalcON}}, until a stream has formed. The misalignment angle is measured in angle-space as the angle between the angle distribution of the stream particles and the frequency vector of the progenitor. We note here that this result is independent of the mass of the progenitor, and the phase of the orbit of the progenitor.

We present the results of this procedure in Table~\ref{StreamActionTable}, where we give the angular difference between the N-body stream structure and the principal eigenvector, $\Delta\varphi$. For all known streams we find that, as with the simulation shown in Fig~\ref{GravNoGrav}, the angle-space distribution rotates by a few degrees. The Anticenter stream exhibits the largest angular change of $\sim 7^\circ$. The misalignment between the streams and the progenitor orbit is still a few degrees, despite the anisotropies in the action-space distribution. Therefore, we expect that orbit-fitting algorithms will not be appropriate for real streams.

\begin{table*}
\caption{Known tidal streams: the approximate actions (given in units of $\ActionUnits$), the ratio of the two largest eigenvalues, $\lambda_1/\lambda_2$, and the misalignment angle, $\varphi$, between the principal eigenvector and the frequency vector at the stream's action coordinates. $\varphi^*$ is the measured misalignment angle from the simulations presented in Section~\ref{Anisotropies}. $k$ and $Q$ give the potential parameters found using an orbit-fitting algorithm on a stream aligned with the principal eigenvector, and $k^*$ and $Q^*$ give the parameters found using an N-body simulation. The true underlying potential has parameters $(k,Q)=(1,1)$. The final column, $\Delta\varphi$, gives the angular difference between the principal eigenvector of the Hessian and the measured direction of the stream from the N-body simulation.}
\begin{tabular}{lllllllllllllllll}
			&&$J_R$	&$|L_z|$	&$J_z$	&&$\frac{\lambda_1}{\lambda_2}$&$\varphi/^\circ$&&$k$&$Q$&&$\varphi^*/^\circ$&$k^*$&$Q^*$&&$\Delta\varphi/^\circ$\\
\hline\\
Anticenter	&&$0.06$&	$3.4$ 	&$0.15$		&&$6$&	$13.0	$&&$0.72$	&$0.15$			&&$6.18$	&$1.01$	&$0.42$ 	&&$7.0$\\
Aquarius	&&$0.34$&	$0.61$	&$0.28$		&&$7$&	$8.0	$&&$2.06$	&$0.10$			&&$7.14$	&$2.20$	&$0.06$ 	&&$2.1$\\
GD-1 		&&$0.29$&	$3.8$ 	&$0.45$		&&$22$&	$3.5	$&&$0.96$	&$0.49$			&&$2.50$	&$0.72$	&$0.50$ 	&&$1.6$\\
NGC 5466	&&$3.4$&	$0.30$	&$2.8$		&&$6$&	$2.8	$&&$0.92$	&$1.05$			&&$1.42$	&$1.10$	&$1.15$ 	&&$2.9$\\
Lethe		&&$0.14$&	$1.2$ 	&$1.3$		&&$29$&	$1.1	$&&$0.90$	&$0.70$			&&$1.97$	&$0.68$	&$1.28$ 	&&$2.3$\\
Cocytos		&&$0.13$&	$0.83$	&$0.99$		&&$28$&	$0.93	$&&$0.82$	&$1.35$			&&$2.18$	&$0.55$	&$0.60$ 	&&$1.5$\\
Palomar 5	&&$0.24$&	$1.2$	&$1.7$		&&$30$&	$0.89	$&&$1.14$	&$1.61$			&&$1.13$	&$0.74$	&$0.75$ 	&&$1.6$\\
Acheron 	&&$0.11$&	$0.50$	&$0.76$		&&$28$&	$0.73	$&&$1.31$	&$0.84$			&&$2.73$	&$1.06$	&$0.35$ 	&&$2.0$\\
Orphan		&&$4.0$&	$5.9$ 	&$0.88$		&&$34$&	$0.64	$&&$1.03$	&$1.10$			&&$0.65$	&$0.72$	&$0.89$  	&&$0.3$\\
Sagittarius	&&$2.3$&	$2.1$ 	&$4.0$		&&$29$&	$0.43	$&&$0.96$	&$1.03$			&&$1.32$	&$1.05$	&$0.92$ 	&&$0.9$\\
Styx 		&&$0.91$&	$0.22$	&$5.6$		&&$37$&	$0.15	$&&$1.01$	&$0.99$			&&$1.17$	&$0.81$	&$0.99$ 	&&$1.3$\\
\hline
\end{tabular}
\label{StreamActionTable}
\end{table*}
\begin{figure}
$$\includegraphics{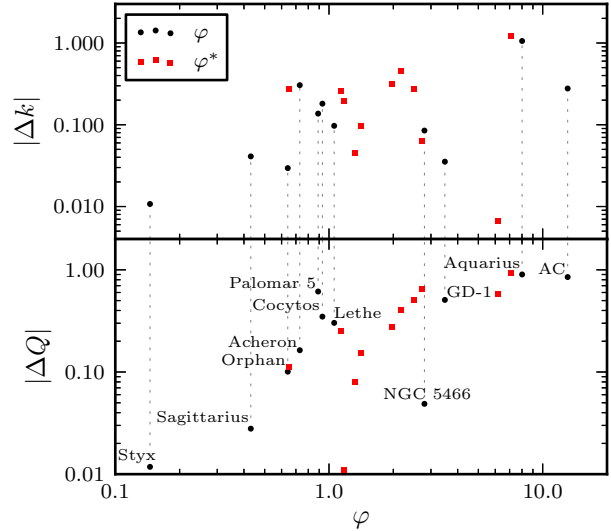}$$
\caption{Errors in the parameters $(k,Q)$ of the two-parameter Galactic potentials obtained when using an orbit-fitting algorithm to analyse known streams plotted against the misalignment angle, $\varphi$, in the true potential with $(k,Q)=(1,1)$. The potential parameters are adjusted until the frequency vector of the progenitor of the stream aligns with angle-space structure of the stream. This simulates the operation of an orbit-fitting algorithm. The round black points show the errors obtained when using an artificial stream perfectly aligned with the principal eigenvector of the Hessian in the true potential. The square red points show the errors when using an N-body simulation. The labels refer to the round black points only, and the light grey lines indicate the correspondence between the round black points in the top and bottom panels.}
\label{DeltaKQ}
\end{figure}

\section{Errors in potential parameters}
Whilst a good indicator of whether an orbit-fitting algorithm is appropriate or not, $\varphi$ does not give a good measure of how much we will err if we use an orbit-fitting algorithm. We would like to know how the magnitude of the misalignment relates to the error in potential parameters found by simply fitting an orbit to the stream. 

We use the suite of two-parameter potentials described in Appendix~\ref{GalaxyPotentials}. These are multi-component Galactic potentials which all have the same circular speed at the Sun, but which vary in two key respects: the flattening of the halo density, $Q$, and the ratio of the magnitude of the force due to the disc and the halo at the Sun, $k$, normalised such that the PJM11 potential is the potential with $(k,Q)=(1,1)$.

For each stream we use the Nelder-Mead algorithm \citep{NelderMead} to adjust $(k,Q)$, until the progenitor frequency vector is aligned with the angle distribution of the stream. The frequency vector is found using an extension of the St\"ackel-fitting algorithm (see appendix of Paper II). This simulates the operation of an orbit-fitting algorithm. The observed stream is misaligned with the orbit but by using an orbit-fitting algorithm we are requiring the `best-fit' potential to make this stream an orbit. This means we need to find a potential where the frequency vector of the stream members is aligned with the stream. This approach neglects the spread in frequencies of the stream members, which one might also want to minimise when orbit-fitting.

We use two stream distributions for each known stream -- one which is aligned with the principal eigenvector of the Hessian in the true potential, and one taken from an N-body simulation which includes the effects of the self-gravity. For the first of these we create a series of $100$ $(\boldsymbol{\theta},\boldsymbol{J})$ points with the same actions as the progenitor, and angles lying at regular intervals along the principal eigenvector of the Hessian. We then use the torus machine to find the corresponding $(\boldsymbol{x},\boldsymbol{v})$ in the true potential $(k,Q)=(1,1)$. For the second approach we use a sample of $100$ particles from each of the simulations given in the previous section.

The results of this experiment are shown in Table~\ref{StreamActionTable}. If the orbit-fitting algorithm is appropriate for a given stream we should recover $(k,Q)=(1,1)$. In Fig.~\ref{DeltaKQ} we plot the errors, $(\Delta k, \Delta Q)$, in the parameters $(k,Q)$ against the misalignment angle for all the streams using both the artificial stream distribution and the N-body distribution. We see that the error in the parameters scales approximately with the misalignment angle, so we expect large misalignment angles lead to large errors in the potential parameters using orbit-fitting algorithms. However, the scatter about this line is reasonably large so the relationship is not simple and other factors are at play.

We begin by discussing the results from the artificial stream distributions (the unstarred values). The Anticenter, Aquarius and GD-1 streams all have $\varphi>3.5^\circ$ and as such have large errors in the potential parameters, particularly the flattening. Notably, for the Aquarius stream the errors are of order one, due to the low actions of the stream. NGC 5466 also has large $\varphi$ but the errors in the potential parameters are $<10\percent$. However, this orbit is awkward to deal with on account of its low $L_z$, yet high $J_R$ and $J_z$. It is in this regime where the largest errors in the actions are expected \citep{Sanders2012}, and correspondingly the largest errors in the frequencies and Hessian. Therefore, the error in $\varphi$ is expected to be large for NGC 5466. Only the Orphan, Sagittarius and Styx streams have small enough misalignment angles that their potential parameters are accurate to $<10\percent$. We therefore expect orbit-fitting algorithms to be appropriate for these streams. The other streams have intermediate potential parameter errors which range from $10$ to $60\percent$, and the use of orbit-fitting algorithms may be appropriate depending on the quality of the data.

From analysing the N-body simulations we find a similar set of results. The Orphan, Styx, Palomar 5 and Sagittarius streams all have small potential parameter errors of $<30\percent$, so orbit-fitting algorithms should be appropriate for these streams. Again NGC 5466 has very small errors of $<15\percent$. For those streams with $\varphi^*\gtrsim2^\circ$ the errors in the parameters are $\gtrsim30\percent$.


We have found that for a realistic Galactic potential, order one errors in the parameters of the potential can arise from naively using an orbit-fitting algorithm on known streams. These results were derived assuming a spherical halo. From the results of the previous section and \citet{EyreBinney2011} we expect that a flattened halo will introduce further error in orbit-fitting algorithms.

\section{Conclusions}

In the next few years more tidal stream data will be collected by surveys of
the Galactic halo, so there is considerable scope for using tidal streams to constrain the Galactic potential at these large scales. However, it is imperative that appropriate algorithms are developed and tested for this end. Here we have provided an in-depth discussion of the applicability of orbit-fitting algorithms which rely on the assumption that a stream delineates an orbit. We have shown that this assumption is necessarily false and can lead to systematic biases.

We presented the angle-action formalism of stream formation, in which
streams form due to their member stars being on different orbits. We
demonstrated that in the angle-action framework streams do not delineate
orbits, and the degree of misalignment depends only on the progenitor orbit, and hence the Galactic potential.

The degree of misalignment was quantified for the logarithmic potential,
which is used in many simulations, and a multi-component realistic Galactic potential. We found that the misalignment angle is small but non-zero for the logarithmic potential. For the realistic Galactic potential we found similar results for orbits which lie far out in the halo, but the misalignment increases significantly as we approach the disc, where the potential flattens. We concluded that tests of orbit-fitting methods which use the logarithmic potential may give unrealistically good results due to its very small misalignment angles.

We have presented a summary of known streams which may be useful for constraining the Galactic potential. For each of these streams we have estimated the actions of the progenitor using data from the literature. At each of these action-space points we quantified the expected misalignment between the stream and the underlying progenitor orbit for a realistic Galactic potential.

Whilst a useful indicator as to whether an orbit-fitting algorithm is appropriate or not, the misalignment angle does not quantify the error involved in estimating the potential parameters from orbit-fitting. We introduced a family of two-parameter realistic Galactic potentials described by the halo-flattening and the halo-to-disc force ratio at the Sun. For each of the known streams we explored this space of potentials until we found the potential which fits an orbit to the stream. As expected the error in the potential parameters is found to correlate approximately with the magnitude of the misalignment angle. We showed that this can introduce order one errors in the potential parameters.

We demonstrated that all these results are essentially independent of the mass of the progenitor up to the mass scale where dynamical friction becomes relevant. Mass acts to scale the angle-action space distributions, whilst leaving the shape unaffected. We therefore expect that even for large progenitor masses, the results are valid. We also showed from N-body simulations that anisotropies in the action-space distribution introduced by the self-gravity of the cluster cause the misalignment of stream particles to change by a few degrees. However, the misalignment for the known streams is still shown to be non-negligible when the effects of self-gravity are included.

The angle-action formalism is a clear framework in which to view and discuss
stream formation. It has enabled us to quantify the errors involved in
orbit-fitting methods for interesting potentials and led to the conclusion
that orbit-fitting algorithms are not appropriate when analysing many streams
in the Milky Way. Hence streams need to be modelled without resort to orbit
fitting, and in the second of these papers \citep{SandersBinney2013} we
present such an alternative algorithm.

\section*{Acknowledgements}
JS acknowledges the support of the Science and Technology Facilities Council and we thank Paul McMillan for useful discussions. We also thank the anonymous referee for their useful suggestions which have improved the paper. 
{\footnotesize{
\bibliographystyle{mn2e-2}
\bibliography{StreamFit}
}}
\appendix
\section{A family of two-parameter potentials}\label{GalaxyPotentials}
We wish to construct a suite of realistic Galaxy potentials which are defined by two parameters: the density flattening of the halo, $Q$, and the ratio $k$ of the force on the Sun due to visible matter and dark matter. This gives us a range of appropriate potentials which we explore to find the best-fit Galaxy potential. It acts as a prior in our exploration of all possible Galaxy potentials.

For our base model we adopt the usual multi-component model: a bulge, thick and thin discs, and a dark halo. For each of these components we use the functional forms discussed by \cite{McMillan2011}. The bulge is taken to be a Bissantz-Gerhard model, and we adopt exponential discs for the thick and thin discs. For the dark matter profile we adopt the NFW profile \citep{NFW1996} with a flattening introduced:
\begin{equation}
\rho_h = \frac{\rho_{\rmn{h},0}}{x(1+x)^2} \mbox{ where } x=\frac{\sqrt{R^2+(z/Q)^2}}{r_h}.
\end{equation}
This introduces the first of our two parameters, the halo flattening $Q$. The second is defined as
\begin{equation}
k\equiv \frac{1}{N}\frac{g_\mathrm{disc}(R_0,z_0)}{g_\mathrm{halo}(R_0,z_0)}
\end{equation}
where $g_i$ is the magnitude of the gravitational force on the Sun due to the $i$th component, and the normalisation $N$ is chosen such that $k=1$ for the `best' potential from \cite{McMillan2011}. The model with $(k,Q)=(1,1)$ corresponds exactly to McMillan's best potential. We take $(R_0,z_0)=(8.29,0.0)\kpc$.

We would like all these potentials to satisfy the observational constraints which have been collected and listed by \cite{McMillan2011}. These include maser observations and terminal velocity curves. However, the most important constraint is the circular speed at the solar position which is largely constrained by the motion of Sgr A* \citep{Reid2004}. Therefore, we only adjust the parameters until the circular speed at the solar position is correct, which \citeauthor{McMillan2011} found to be $v_c=239.1\kms$. 

\subsection{Procedure}
For a given pair of the parameters $(k_s,Q_s)$ we follow this procedure to find a realistic Galactic potential with these parameters:
\begin{enumerate}
\item Construct McMillan's best potential corresponding to $(k,Q)=(1,1)$.
\item Set $Q=Q_s$.
\item Adjust $\rho_{\rmn{h},0}$ until $k=k_s$.
\item Calculate the circular speed at the solar position in this model, $v_{cs}$. 
\item Scale $\rho_{\rmn{h},0}$ and $\Sigma_{\rmn{d}}$ by the same factor $p = (v_c/v_{cs})^2$ so that the circular speed at the solar position is restored to $v_c=239.1\kms$.
\end{enumerate}

\subsection{Tabulation}
\begin{figure}
$$\includegraphics{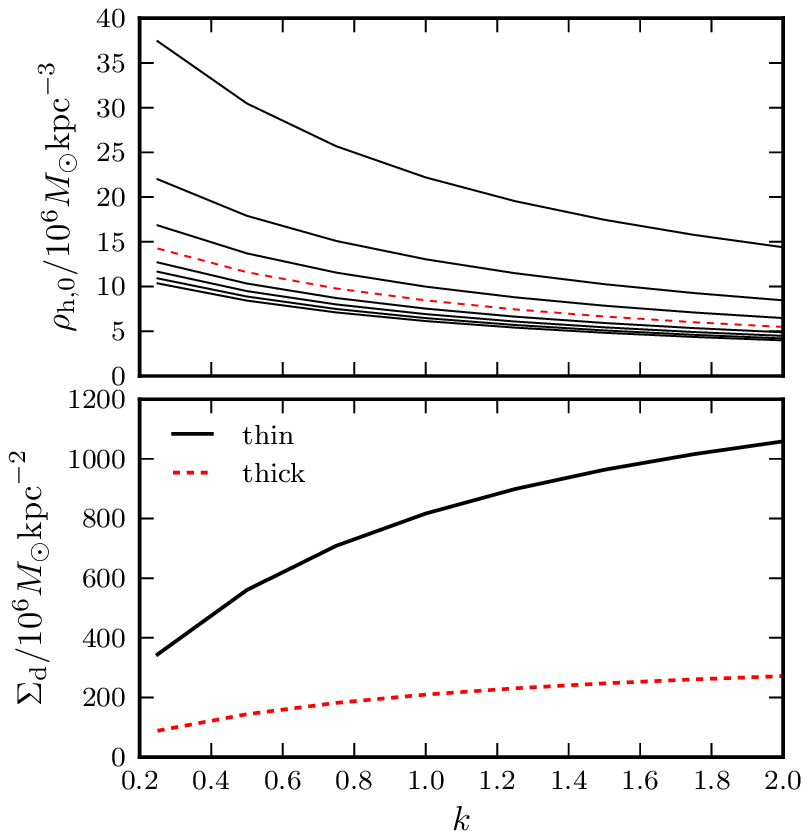}$$
\caption{$\rho_{\rmn{h},0}$ and $\Sigma_{\rmn{d}}$ as a function of $k$ for the two-parameter $(k,Q)$ potentials. The top panel shows lines of constant $Q$. The lines are spaced by $\Delta Q=0.25$ with the uppermost line showing $Q=0.25$. The red dashed line gives $Q=1$. The bottom panel shows the variation of $\Sigma_{\rmn{d}}$ for the thin and thick discs, which has no $Q$ dependence.}
\label{kQplot}
\end{figure}

As the constructed potentials are simply described by two parameters it is convenient to construct a 2D grid of these potentials in $(k,Q)$ space. We construct the potentials using the above procedure for $N$ values of $k$ and $N$ values of $Q$. At each of these points in parameter space we store $\rho_{\rmn{h},0}$ and $\Sigma_{\rmn{d}}$ in an $N\times N$ array. This grid may then be linearly interpolated for a given pair of $(k,Q)$. For any call which falls outside the grid range we use the full procedure outlined above. In Table~\ref{kQtable} we list the parameters of a sample of $(k,Q)$ models, and in Fig.~\ref{kQplot} these parameters are plotted as a function of $k$.

\subsection{Other Uses}
We have introduced this family of potentials to construct a simple, yet realistic, space of potentials which we can explore when applying the stream-fitting algorithm. This set of potentials should also be useful for many other tasks. It provides a space of potentials which satisfy experimental constraints which can be explored when applying any potential fitting algorithm. As there are only two parameters it is fast and simple to calculate and communicate the results, and could be used as a basis for testing potential fitting routines against each other. The reduction of the number of parameters down to a minimal set, which still includes all the necessary complexities, is clearly advantageous.

The two parameters describe the large scale structure of the Galaxy simply. Large scale features, such as the shape and visible-dark matter ratio, are important when trying to compare large-scale galaxy evolution models with the Galaxy. This family of potentials may be useful in this respect to provide a simple way to compare and discuss the results of simulations.

\begin{table*}
\caption{Parameters for 64 $(k,Q)$ models: $\rho_{\rmn{h},0}$ is the central dark matter halo density (in units of $10^6M_\odot\kpc^{-3}$) and $\Sigma_{\rmn{d},\rmn{thin}}$ and $\Sigma_{\rmn{d},\rmn{thick}}$ are the surface densities of the thin and thick disc (in units of $10^6M_\odot\kpc^{-2}$). The top lines with $(k,Q)=(1,1)$ are identical to McMillan's best potential.}
\begin{minipage}{\columnwidth}
\begin{tabular}{lllll}
$k$&$Q$&$\rho_{\rmn{h},0}$&$\Sigma_{\rmn{d},\rmn{thin}}$&$\Sigma_{\rmn{d},\rmn{thick}}$\\
\hline
$1$&$1$&$8.46$&$817$&$209$\\
\hline
$0.25$&$0.25$&$37.4$&$344$&$88.3$\\
$0.25$&$0.50$&$22.0$&$344$&$88.3$\\
$0.25$&$0.75$&$16.8$&$344$&$88.3$\\
$0.25$&$1.00$&$14.3$&$344$&$88.3$\\
$0.25$&$1.25$&$12.7$&$344$&$88.3$\\
$0.25$&$1.50$&$11.7$&$344$&$88.3$\\
$0.25$&$1.75$&$10.9$&$344$&$88.3$\\
$0.25$&$2.00$&$10.4$&$344$&$88.3$\\
\hline
$0.50$&$0.25$&$30.5$&$560$&$144$\\
$0.50$&$0.50$&$17.9$&$560$&$144$\\
$0.50$&$0.75$&$13.7$&$560$&$144$\\
$0.50$&$1.00$&$11.6$&$560$&$144$\\
$0.50$&$1.25$&$10.3$&$560$&$144$\\
$0.50$&$1.50$&$9.49$&$560$&$144$\\
$0.50$&$1.75$&$8.88$&$560$&$144$\\
$0.50$&$2.00$&$8.43$&$560$&$144$\\
\hline
$0.75$&$0.25$&$25.7$&$709$&$182$\\
$0.75$&$0.50$&$15.1$&$709$&$182$\\
$0.75$&$0.75$&$11.6$&$709$&$182$\\
$0.75$&$1.00$&$9.78$&$709$&$182$\\
$0.75$&$1.25$&$8.71$&$709$&$182$\\
$0.75$&$1.50$&$8.00$&$709$&$182$\\
$0.75$&$1.75$&$7.49$&$709$&$182$\\
$0.75$&$2.00$&$7.10$&$709$&$182$\\
\hline
$1.00$&$0.25$&$22.2$&$817$&$209$\\
$1.00$&$0.50$&$13.0$&$817$&$209$\\
$1.00$&$0.75$&$9.99$&$817$&$209$\\
$1.00$&$1.00$&$8.46$&$817$&$209$\\
$1.00$&$1.25$&$7.53$&$817$&$209$\\
$1.00$&$1.50$&$6.92$&$817$&$209$\\
$1.00$&$1.75$&$6.47$&$817$&$209$\\
$1.00$&$2.00$&$6.14$&$817$&$209$\\
\hline
\end{tabular}
\end{minipage}
\begin{minipage}{\columnwidth}
\begin{tabular}{lllll}
$k$&$Q$&$\rho_{\rmn{h},0}$&$\Sigma_{\rmn{d},\rmn{thin}}$&$\Sigma_{\rmn{d},\rmn{thick}}$\\
\hline
$1$&$1$&$8.46$&$817$&$209$\\
\hline
$1.25$&$0.25$&$19.5$&$899$&$231$\\
$1.25$&$0.50$&$11.5$&$899$&$231$\\
$1.25$&$0.75$&$8.80$&$899$&$231$\\
$1.25$&$1.00$&$7.45$&$899$&$231$\\
$1.25$&$1.25$&$6.63$&$899$&$231$\\
$1.25$&$1.50$&$6.09$&$899$&$231$\\
$1.25$&$1.75$&$5.70$&$899$&$231$\\
$1.25$&$2.00$&$5.41$&$899$&$231$\\
\hline
$1.50$&$0.25$&$17.5$&$964$&$247$\\
$1.50$&$0.50$&$10.3$&$964$&$247$\\
$1.50$&$0.75$&$7.86$&$964$&$247$\\
$1.50$&$1.00$&$6.65$&$964$&$247$\\
$1.50$&$1.25$&$5.93$&$964$&$247$\\
$1.50$&$1.50$&$5.44$&$964$&$247$\\
$1.50$&$1.75$&$5.09$&$964$&$247$\\
$1.50$&$2.00$&$4.83$&$964$&$247$\\
\hline
$1.75$&$0.25$&$15.8$&$1020$&$261$\\
$1.75$&$0.50$&$9.27$&$1020$&$261$\\
$1.75$&$0.75$&$7.10$&$1020$&$261$\\
$1.75$&$1.00$&$6.01$&$1020$&$261$\\
$1.75$&$1.25$&$5.35$&$1020$&$261$\\
$1.75$&$1.50$&$4.92$&$1020$&$261$\\
$1.75$&$1.75$&$4.60$&$1020$&$261$\\
$1.75$&$2.00$&$4.36$&$1020$&$261$\\
\hline
$2.00$&$0.25$&$14.4$&$1060$&$272$\\
$2.00$&$0.50$&$8.46$&$1060$&$272$\\
$2.00$&$0.75$&$6.48$&$1060$&$272$\\
$2.00$&$1.00$&$5.48$&$1060$&$272$\\
$2.00$&$1.25$&$4.88$&$1060$&$272$\\
$2.00$&$1.50$&$4.48$&$1060$&$272$\\
$2.00$&$1.75$&$4.20$&$1060$&$272$\\
$2.00$&$2.00$&$3.98$&$1060$&$272$\\
\hline
\end{tabular}
\end{minipage}
\label{kQtable}
\end{table*}

\bsp

\label{lastpage}

\end{document}